\pdfoutput=1  
\documentclass[11pt,a4paper]{article}

\usepackage[T1]{fontenc}
\usepackage[utf8]{inputenc}
\usepackage{lmodern}
\usepackage[protrusion=true,expansion=true,final]{microtype}
\usepackage[a4paper,top=1in,bottom=1in,left=1.1in,right=1.1in]{geometry}

\usepackage{amsmath}
\usepackage{amssymb}
\usepackage{amsfonts}
\usepackage{bm}
\usepackage{booktabs}
\usepackage{array}
\usepackage{tabularx}
\newcolumntype{Y}{>{\raggedright\arraybackslash}X}
\usepackage{graphicx}
\usepackage{xcolor}
\usepackage{enumitem}
\usepackage{float}
\usepackage{url}
\expandafter\def\expandafter\UrlBreaks\expandafter{\UrlBreaks\do\-\do\_}
\usepackage[round,authoryear]{natbib}
\usepackage[font=small,labelfont=bf]{caption}

\usepackage[colorlinks=true,linkcolor=blue!50!black,citecolor=blue!50!black,urlcolor=blue!50!black]{hyperref}

\setlist{noitemsep,topsep=2pt,parsep=0pt,partopsep=0pt}
\graphicspath{{figures/}{./}}

\title{\textbf{Forgetting Is Not a Fix:}\\[2pt]
\textbf{Path Dependence in Sequential Engram Editing}}
\author{Ferdinand M. Schessl\\\small\texttt{contact@typx.org}}
\date{\today}

\begin{document}
\maketitle

\section{Abstract}

AI Engram \citep{kwon2026engram} formalizes the four engram criteria of neuroscience as a constrained inverse problem in weight space and solves it closed-form: concept-specific memory traces become linear objects that can be extracted once and combined arithmetically. Appendix~F of that work states the Compositional Memory States Hypothesis: edited models live on ``a commutative manifold where the integration of A and B reaches a consistent equilibrium regardless of the learning sequence.'' The paper's evidence base consists of single and paired edits --- in materials terms, single-cycle tests. Fatigue accumulation is structurally invisible in a single-cycle test; whether the hypothesis holds under sequential load is exactly the ``temporal dynamics'' question the paper defers to future work. We run that test on the authors' own reference implementation, at the edit strength the authors themselves selected for the TOFU LLM benchmark (fixed $\alpha = 0.6$, from their grid-search range $\{0.05\text{--}1.0\}$), with pre-registered predictions, across three material charges (Qwen3-0.6B, TinyLlama-1.1B-Chat, Qwen2-0.5B-Instruct --- two vendors, two architecture families). Four findings replicate across all three charges. (1)~Zero-shot composition and sequential re-calibrated editing diverge by 61--71\% of the total edit magnitude (relative Frobenius distance over edited layers). (2)~Cut order is not interchangeable, and the effect is overlap-ordered: high-overlap concept pairs produce order differences up to the scale of an entire edit (weight-distance ratio 0.81--1.00 vs.\ 0.47--0.68 for distant pairs); in one charge the order of cutting two Paris landmarks decides whether an uninvolved third concept survives (NLL 0.85 vs.\ 6.50 nats). (3)~The layer-input covariances of surviving concepts --- the method's own sufficient statistics, read here as strain gauges --- drift monotonically with every further cut, in every surviving concept, in every charge: load-history accumulates. (4)~The forgotten state is not a fixed point of the artifact's future: subsequent cuts partially restore erased knowledge (most pronounced in the smallest charge, NLL $8.9 \to 4.9$ nats against a $0.05$ baseline). The proximity structure of collateral damage (their finding C.4) replicates cleanly only where capacity reserve exists; at $2.4\times$ relative edit magnitude it breaks into broadband collateral --- itself an informative boundary condition. Appendix~F's commutative-manifold hypothesis is thereby falsified for sequential editing in all three charges; the single-edit results of the original paper are untouched. For unlearning-as-compliance the consequence is concrete: erasure certified today does not certify the artifact after its next edit.

\vspace{4pt}
\noindent\textbf{Keywords:} model editing, machine unlearning, engram, path dependence, non-commutativity, materials-science framing, sequential editing, superposition.

\section{1. Introduction}

\citet{kwon2026engram} (ICML 2026 oral) is a methodologically strong formalization of memory localization in neural networks: the four engram criteria (specificity, reactivation, sufficiency, necessity) become a constrained inverse problem, solved closed-form as a data-dependent spectral projector $W^{+} = W \Sigma^{+} (\Sigma^{+} + \Sigma^{-})^{\dagger}$; Theorem~6.1 identifies this with the minimum-norm projection under a Fisher metric, and Corollary~6.3 identifies ablation with a single natural-gradient step on the forgetting loss. Editing becomes linear arithmetic on extracted engrams, in the lineage of task arithmetic \citep{ilharco2023task} --- combinatorial unlearning without the iterative optimization that mass-editing methods such as MEMIT \citep{meng2022memit} require.

The paper is explicit about its epistemic position: identification is ``functional \ldots not learning-trajectory reconstruction,'' retrospective-only (Appendix~H). Its Appendix~F then states a hypothesis that goes beyond the retrospective frame: compositional memory states form ``a commutative manifold where the integration of A and B reaches a consistent equilibrium regardless of the learning sequence,'' with $2^{n}-1$ states reachable by zero-shot arithmetic. The evidence offered for this hypothesis consists of single edits and pairwise compositions, each measured once against a fixed reference state. No sequential, repeated, or order-swapped editing of the same model instance is reported. The conclusion names the gap itself: ``Future work may examine the temporal dynamics of engrams in continual learning.''

This note supplies that examination in its minimal form. The framing is materials testing: a single load cycle cannot exhibit fatigue accumulation, damage-path dependence, or load-history effects --- not because they are absent, but because a one-cycle protocol has no axis on which they could appear. Whether an editing method's linearity survives \emph{sequential} load is therefore not answerable from single-edit evidence, however clean. It requires cutting the same specimen repeatedly and instrumenting the survivors.

\vspace{2pt}
\noindent\textbf{Contributions.}
\begin{enumerate}
\item A sequential two-arm protocol on the authors' own reference implementation and edit regime (their package, their grid-search-selected $\alpha = 0.6$ on the TOFU LLM benchmark, their default scaling), with four internally pre-registered predictions (V1--V4) fixed before measurement.
\item Replication across three material charges spanning two vendors, two architecture families, and 0.5--1.1B parameters, with per-charge raw data published.
\item Falsification of the Appendix-F hypothesis for sequential editing in all three charges: path dependence (V1), overlap-ordered non-commutativity (V2), and monotonically accumulating drift of the survivors' sufficient statistics (V4) --- measured with the method's own instruments.
\item Two boundary findings with practical consequence: collateral proximity structure (their C.4) requires capacity reserve (V3 breaks at $2.4\times$ relative edit magnitude), and erased knowledge partially returns under subsequent cuts --- the forgotten state is not a fixed point.
\end{enumerate}

\noindent\emph{Scope note, stated early:} nothing here contests the single-edit results, the closed-form estimator, or the engram-identification framework of the original paper. What is falsified is one hypothesis (Appendix~F) about the \emph{composition} of edits, in the regime the hypothesis itself claims: sequences. A null result --- sequential arms landing on the same manifold --- was pre-registered as equally reportable support for Appendix~F. That is not what we measured.

\section{2. Materials Framing}

We treat sequential editing as a fatigue protocol on a weight-space specimen:

\begin{table}[H]
\centering
\small
\begin{tabularx}{\linewidth}{@{}l Y@{}}
\toprule
Materials testing & This experiment \\
\midrule
Test specimen & one copy of the base model under an ordered cut sequence \\
Material charge & base model (Qwen3-0.6B / TinyLlama-1.1B-Chat / Qwen2-0.5B-Instruct) \\
Load cycle & one engram ablation (``cut'') at $\alpha = 0.6$ \\
Load path & cut sequence A$\to$C$\to$E; order-swapped pairs for commutativity \\
Damage indicator & teacher-forced NLL (nats) on a fixed probe per concept \\
Strain gauge on survivors & layer-input covariance of each surviving concept's forget-set (the method's own sufficient statistics), relative Frobenius drift vs.\ the virgin state \\
Residual-stress redistribution & collateral NLL shifts in uncut concepts \\
Fatigue accumulation & monotone growth of survivor covariance drift over successive cuts \\
\bottomrule
\end{tabularx}
\end{table}

Two arms operationalize the difference between the paper's composition claim and sequential reality:

\begin{itemize}
\item \textbf{Arm 1 (zero-shot composition, their Appendix-F reading):} all engrams are extracted once on the virgin model $M_0$ and applied in sequence as linear arithmetic. This is the commutative-manifold operation: extraction ignores load history by construction.
\item \textbf{Arm 2 (re-calibrated sequence):} after each cut, the next engram is extracted on the \emph{current} model state. This is what sequential deployment --- compliance-driven deletion requests arriving over time --- actually does.
\end{itemize}

If Appendix~F holds, the two arms land on the same manifold point (weight distance $\approx 0$ at matched cut sets), cut order is interchangeable, and survivor statistics stay put. Each prediction below was fixed before measurement (internally pre-registered 2026-07-05; the verbatim design and a deviations section are in the companion repository's \texttt{PRE\_REGISTRATION.md}).

\begin{itemize}
\item \textbf{V1 (path dependence):} Arm~1 $\neq$ Arm~2; weight distance between arms $\gg 0$ on the scale of the total edit magnitude.
\item \textbf{V2 (non-commutativity, overlap-ordered):} $\mathrm{dist}(M_{AB}, M_{BA}) > \mathrm{dist}(M_{AF}, M_{FA})$ --- the order effect is larger for a high-overlap pair (Eiffel/Louvre, shared ``Paris'' context) than for a distant pair (Eiffel/Python).
\item \textbf{V3 (collateral proximity structure):} after cutting A, collateral NLL rise is proximity-ordered (their C.4, transplanted to sequence step~1).
\item \textbf{V4 (survivor drift):} covariance drift of uncut concepts is (a)~$> 0$, (b)~monotonically accumulating over cuts, (c)~proximity-ordered.
\end{itemize}

\section{3. Protocol}

\begin{sloppypar}
\noindent\textbf{Package and regime.} \texttt{ai-engram} (the authors' public reference implementation), applied unmodified via its public API (\texttt{get\_engram}, \texttt{apply\_engram}, \texttt{EngramEditor.collect\_statistics}). Edit strength $\alpha = 0.6$ --- the value the authors selected for their fixed-$\alpha$ Engram variant on the TOFU LLM unlearning benchmark \citep{maini2024tofu} (grid search over $\{0.05\text{--}1.0\}$; their best fixed-$\alpha$ setting on exact-memorization unlearning). We adopt their own value rather than tuning $\alpha$ to amplify the effect; a smoke test at $\alpha = 1.0$ saturated collateral across the board. Scaling stays at the package default (count ratio $n/N$). All-layer editing, FP32, CPU, deterministic (no sampling anywhere in the pipeline).
\end{sloppypar}

\noindent\textbf{Concept set.} Six concepts with graded semantic overlap: A~Eiffel Tower--Paris, B~Louvre--Paris (high overlap with A), C~Colosseum--Rome (same category, different city), D~Brandenburg Gate--Berlin, E~insulin--pancreas (distant domain), F~Python--van Rossum (distant domain). Per concept: four forget sentences (extraction set) and one held-out cloze probe. Ten neutral retain sentences complete the total-set context for extraction. Damage indicator: teacher-forced mean NLL over the probe's answer tokens.

\noindent\textbf{Sequence.} Cuts A$\to$C$\to$E on one specimen per arm. After each Arm-2 cut, survivor covariance drift is measured as the mean relative Frobenius change of the layer-input covariances of each remaining concept's forget-set against the virgin state. V1 compares final Arm-1 and Arm-2 weights (mean relative Frobenius distance over the union of edited layers), scaled by $\mathrm{dist}(M_0, \mathrm{Arm\,1})$. V2 runs both cut orders for the overlap pair (A,B) and the distant pair (A,F) on fresh copies, Arm-2 style, reporting the weight-distance ratio (order difference / single-composition magnitude) and the maximum per-concept NLL difference between orders.

\noindent\textbf{Charges.} Qwen3-0.6B \citep{qwen3} (their quickstart model), TinyLlama-1.1B-Chat-v1.0 \citep{tinyllama} (Llama architecture, different vendor and training provenance), Qwen2-0.5B-Instruct \citep{qwen2} (earlier generation, smallest capacity). A fourth charge (Qwen2-1.5B-Instruct) died twice with kernel OOM kills during Arm~1 on a 128-GB host ($8960^2$ MLP covariances $\times$2 statistics sets $+$ SVD workspace) --- an infrastructure failure documented in the logs, not an empirical finding. Llama-3.2-1B (their TOFU substrate) is gated and was not accessible; noted as a limitation.

\noindent\textbf{Runtimes.} 2\,222\,s / 15\,618\,s / 5\,011\,s per charge (CPU); everything reproduces from one script (\texttt{seq\_c4\_test.py --model <id>}).

\section{4. Results}

Consolidated across the three charges (per-charge raw JSONs published; \checkmark~$=$ prediction met):

\begin{table}[H]
\centering
\footnotesize
\renewcommand{\arraystretch}{1.25}
\begin{tabularx}{\linewidth}{@{}p{2.5cm} Y Y Y l@{}}
\toprule
Prediction & Qwen3-0.6B & TinyLlama-1.1B & Qwen2-0.5B & Total \\
\midrule
\textbf{V1} arm divergence / edit magnitude
& \textbf{0.710} \checkmark & \textbf{0.642} \checkmark & \textbf{0.614} \checkmark & \textbf{3/3}, narrow band \\
\textbf{V2} order effect: overlap vs.\ distant (ratio; max~$|\Delta$NLL$|$)
& \textbf{0.997} (5.65) vs.\ 0.576 (1.63) \checkmark
& \textbf{0.912} (2.48) vs.\ 0.466 (1.15) \checkmark
& \textbf{0.809} (9.57) vs.\ 0.677 (5.99) \checkmark
& \textbf{3/3} \\
\textbf{V3} collateral proximity-ordered after cut~A
& B~$+1.57 \gg$ D~$+0.76 >$ C/F~$+0.08 >$ E~$\approx 0$ \checkmark
& B~$+0.96$, rest~$\approx 0$ (\checkmark)
& \textbf{FAIL}: C~$+8.5 >$ D/F~$+5.5 >$ B~$+3.8$
& \textbf{2/3} \\
\textbf{V4a/b} survivor drift $> 0$, monotone accumulating
& \checkmark (B $1.64\!\to\!1.80$) & \checkmark (B $0.41\!\to\!0.79$) & \checkmark (B $1.24\!\to\!2.45$)
& \textbf{3/3} \\
\textbf{V4c} drift proximity-ordered
& \checkmark (B $\gg$ C/D $\gg$ E/F) & \checkmark & (\checkmark) weak & 2.5/3 \\
\textbf{Bonus} erased state not a fixed point
& A $13.20\!\to\!9.88$ & weak ($\pm 0.3$) & \textbf{A $8.91\!\to\!5.21\!\to\!4.85$} & replicated \\
\bottomrule
\end{tabularx}
\end{table}

\begin{figure}[H]
\centering
\includegraphics[width=\linewidth]{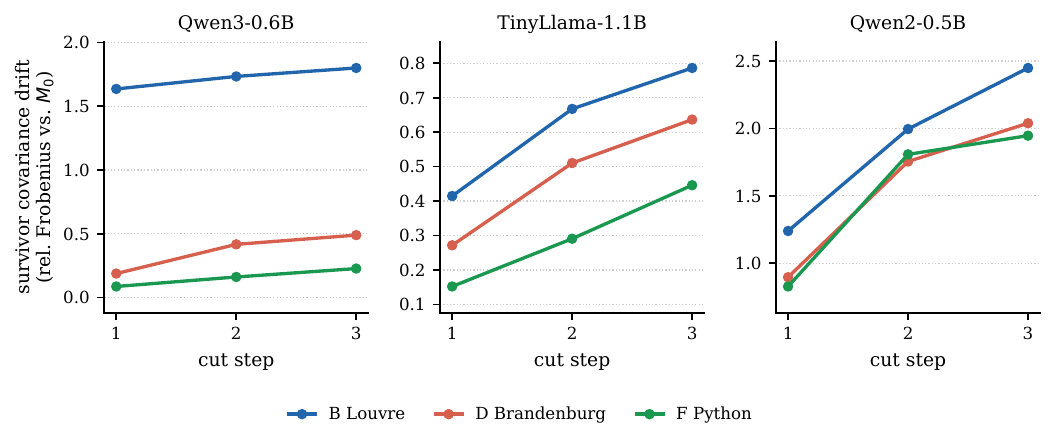}
\caption{\textbf{V4 --- the survivors keep a ledger.} Relative Frobenius drift of the layer-input covariances (the method's own sufficient statistics) for the three concepts that survive all cuts, measured on concepts that were never touched. Monotone growth over the cut sequence in every surviving concept, in every charge (twelve series total, zero exceptions). Note the differing $y$-scales: the accumulation \emph{pattern}, not its absolute size, is the invariant.}
\label{fig:drift}
\end{figure}

\begin{figure}[H]
\centering
\includegraphics[width=\linewidth]{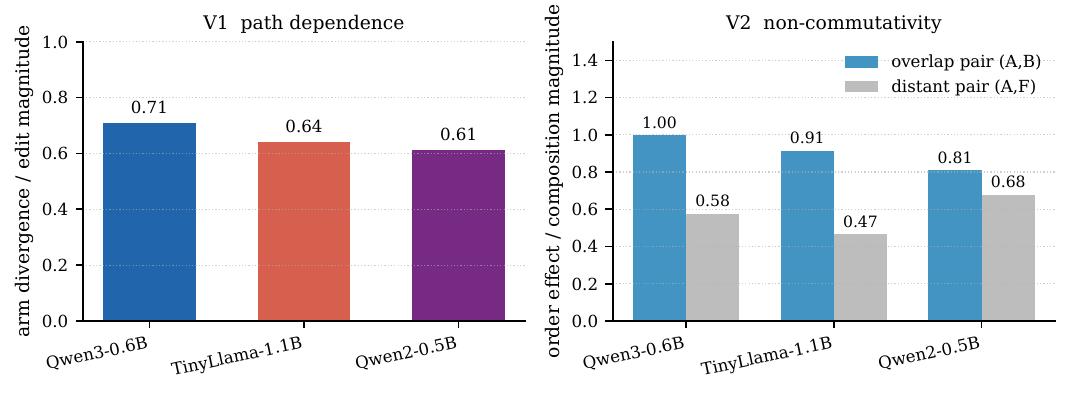}
\caption{\textbf{V1 path dependence and V2 non-commutativity.} Left: the re-calibrated sequence (Arm~2) diverges from the zero-shot composition (Arm~1) by 61--71\% of the total edit magnitude, in a narrow band across all three charges. Right: the order effect (weight distance between the two cut orders, over the single-composition magnitude) is larger for the semantically overlapping pair than for the distant pair in every charge.}
\label{fig:pathdep}
\end{figure}

\noindent\textbf{V1 --- zero-shot composition and sequential reality are different objects.} The two arms end 61--71\% of the total edit magnitude apart (Fig.~\ref{fig:pathdep}, left) --- a narrow band across two vendors, two architecture families, and a $2.2\times$ size range. Behaviorally the arms disagree about what the edit \emph{cost}: in charge~1, Louvre collateral is 2.82 nats (Arm~1) vs.\ 6.93 nats (Arm~2). Whatever quantity a zero-shot composition certifies, it is not the state a sequentially edited artifact is actually in.

\noindent\textbf{V2 --- cut order matters, and semantic overlap decides how much.} For the high-overlap pair, the two orders differ by 81--100\% of a full composition's magnitude (Fig.~\ref{fig:pathdep}, right) --- cutting A then B and cutting B then A produce models as far apart as the edit itself is large. Distant pairs stay at 47--68\%. The sharpest single observation (charge~1): the order of cutting Eiffel and Louvre decides the fate of the \emph{uninvolved} Colosseum --- NLL 0.85 nats in one order, 6.50 in the other. Redistribution routes the damage, and the routing reads the load path.

\noindent\textbf{V4 --- the survivors keep a ledger.} Survivor covariance drift is positive after the first cut, grows with every further cut, in every surviving concept, in every charge (Fig.~\ref{fig:drift}; twelve series --- nine with three load points, three with two --- zero exceptions). In materials terms: each cycle leaves the neighbors more deformed; the method's own sufficient statistics --- collected on concepts that were never touched --- record the specimen's load history. This is fatigue accumulation, measured with the instrument the method itself ships.

\noindent\textbf{V3 --- proximity structure needs capacity reserve.} Charge~1 reproduces C.4's proximity ordering cleanly at sequence step~1. Charge~2 shows only the top contrast (its concept anchoring is weak to begin with: probe baselines 5.7--6.3 nats --- the model barely holds the facts under load at all). Charge~3 breaks the ordering: collateral is broadband (even insulin rises 13 nats, $2.2 \to 15.2$, after the Colosseum cut in Arm~1). The candidate explanation is measurable: the same $\alpha$ produces a $2.4\times$ larger relative edit magnitude in the 0.5B charge ($\mathrm{dist}(M_0, \mathrm{Arm\,1}) = 0.81$ vs.\ $0.34/0.31$). Two readings remain open --- denser superposition at scarce capacity \citep{elhage2022superposition} smearing the collateral topology, or NLL saturation ($> 8$ nats) drowning the ordering. Deciding between them requires magnitude-calibrated $\alpha$ per charge (equal $\mathrm{dist}(M_0, M_1)$), which is the designated follow-up. Until then the honest statement is: proximity-structured collateral is confirmed only with capacity reserve.

\section{5. Forgetting Is Not a Fix}

\begin{figure}[H]
\centering
\includegraphics[width=0.62\linewidth]{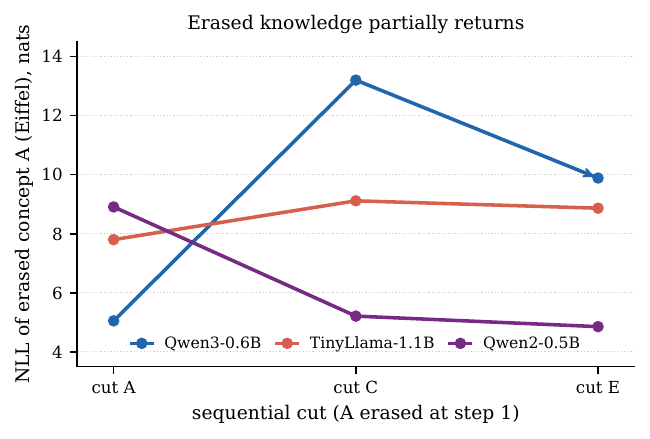}
\caption{\textbf{Erased knowledge partially returns.} NLL of concept A (Eiffel), erased at cut step~1, as two \emph{unrelated} concepts (C, then E) are subsequently cut in Arm~2. In the two capacity-bearing charges the erased NLL falls back toward baseline --- monotonically in Qwen2-0.5B ($8.91 \to 5.21 \to 4.85$), after an intervening rise in Qwen3-0.6B. TinyLlama-1.1B is the weak charge (probe baseline $\approx 6$ nats: the fact is barely held even before editing). Virgin baselines: 0.09 / 6.03 / 0.05 nats.}
\label{fig:return}
\end{figure}

The bonus finding deserves its own section because it carries the unlearning consequence. In charges~1 and 3, knowledge that a cut had pushed to high NLL returns partially when \emph{other} concepts are subsequently cut (Fig.~\ref{fig:return}): charge~1, Eiffel $13.20 \to 9.88$ nats after the insulin cut; charge~3, Eiffel $8.91 \to 5.21 \to 4.85$ nats across the two subsequent cuts, against a virgin baseline of $0.05$. Redistribution does not respect the forgetting target: the same coupling that routes collateral damage \emph{into} neighbors can route function \emph{back into} the erased region.

For unlearning-as-compliance (right-to-be-forgotten workflows, safety-motivated removal) this inverts a quiet assumption: that erasure, once verified, stays verified. On this evidence, an erasure certificate is a statement about a \emph{state}, and the state is not a fixed point of the artifact's maintenance process. A model that passes the deletion audit today can partially restore the deleted association after its next routine edit --- without anyone touching the deleted concept. Verification must therefore be re-run after \emph{every} subsequent edit, or the editing process itself must come with a stability guarantee under composition. Appendix~F was that guarantee-candidate; it is the hypothesis these measurements falsify.

\section{6. What This Note Does Not Show}

\begin{itemize}
\item \textbf{No contest of the single-edit results.} Estimator, engram identification, and the single-cycle experiments of \citet{kwon2026engram} stand untouched. Corollary~6.3 (ablation $=$ one natural-gradient step) is \emph{consistent} with everything here: one step is the elastic regime; our findings begin at step two.
\item \textbf{Three small charges, one concept set.} 0.5--1.1B parameters, six English factual concepts, one cut sequence plus two order-swapped pairs. The TOFU substrate itself (Llama-3.2-1B) was gated. Effect \emph{presence} replicated 3/3 with tight V1 ratios; effect \emph{sizes} elsewhere (notably V2 max $\Delta$NLL) vary by charge.
\item \textbf{One global $\alpha$, not swept.} We test the authors' own selected value ($0.6$); it is not magnitude-calibrated per charge (the V3 break shows what that costs), nor swept toward the small-$\alpha$ limit, where first-order composition should be most accurate and Corollary~6.3's single-step equivalence lives. The claim is about their operating point, not all $\alpha$; the magnitude-calibrated re-run is specified, not executed.
\item \textbf{Self-chosen distance scales.} Relative Frobenius means over edited layers are one reasonable convention; ratios, not the absolute numbers, carry the claims.
\item \textbf{No recovery cycles.} The pre-registered optional arm (brief retain-fine-tuning after each cut: does $\Sigma$ return elastically or drift plastically?) was not run.
\item \textbf{Deterministic point measurements.} No seeds, no CIs; the design trades statistical apparatus for exact reproducibility (every number regenerates bit-identically from one script call).
\item \textbf{The materials vocabulary is a measurement frame, not a claimed mechanism.} ``Fatigue'' here names a measured monotone accumulation pattern in sequential edits, not a claim that transformer weights and metal alloys share physics. The frame earns its keep by having specified V1--V4 --- including the informative-null reading --- before measurement.
\end{itemize}

\section{7. Conclusion}

The original paper closes: ``Future work may examine the temporal dynamics of engrams in continual learning.'' Examined, in the smallest honest version: the temporal dynamics are path-dependent, non-commutative in proportion to concept overlap, and cumulative in the survivors' own sufficient statistics --- in all three charges, on the authors' implementation, at the edit strength they selected for LLM unlearning. The commutative manifold of Appendix~F does not survive its first sequence. What replaces it is a materials picture: every cut redistributes load, the redistribution reads the load path, and the specimen keeps the ledger. An edited model is not a point on a state manifold reachable in any order; it is an artifact with a history.

\vspace{4pt}
\noindent\textbf{Reproducibility.} Companion repository \url{https://github.com/FerdinandSchessl/engram-seq-note-companion} --- the experiment script (\texttt{seq\_c4\_test.py}, one script, \texttt{--model} switch, pinned \texttt{ai-engram==0.8.0}), the three raw-result JSONs, the verbatim pre-registration with a deviations section, and \texttt{verify\_numbers.py}, which re-derives every number quoted in this note from the raw JSONs (57 checks).

\bibliographystyle{plainnat}
\bibliography{refs}

\end{document}